\documentclass{aa}  

\usepackage{graphicx}
\usepackage{txfonts}                                 
\usepackage{placeins}  
\usepackage{multirow}
\usepackage{subcaption}
\usepackage{hyperref}
\hypersetup{
    colorlinks=true,
    linkcolor=blue,
    citecolor=blue,
    urlcolor=blue
}

\begin{document}

   \title{Galaxy and Halo Assembly Bias in Alternative Dark Matter Models}

   \author{
Yikun Wang\inst{1}\corrauth{yxw2032@case.edu}
\and
Idit Zehavi\inst{1}\email{idit.zehavi@case.edu}
\and
Sergio Contreras\inst{2}\email{el.hantke@gmail.com}
\and
Jonás Chaves-Montero\inst{3}\email{jchaves@ifae.es}
\and
Giulia Despali\inst{4,5,6}\email{giulia.despali@unibo.it}
\and
Carlo Giocoli\inst{5,6}\email{carlo.giocoli@unibo.it}
\and
Lauro Moscardini\inst{4,5,6}\email{lauro.moscardini@unibo.it}
\and
Massimiliano Romanello\inst{4,5}\email{massimilia.romanell2@unibo.it}
\and
Mark Vogelsberger\inst{7,8}\email{mvogelsb@mit.edu}
}

\institute{
Department of Physics, Case Western Reserve University,
10900 Euclid Avenue, Cleveland, OH 44106-1715, USA
\and
Facultad de F\'isica, Universidad de Sevilla, Campus de Reina Mercedes, Av. Reina Mercedes s/n 41012 Seville, Spain
\and
Institut de Física d'Altes Energies,
Campus UAB, Facultat Ciencies Nord,
08193 Bellaterra, Barcelona, Spain
\and
Dipartimento di Fisica e Astronomia ``Augusto Righi'',
Alma Mater Studiorum Università di Bologna,
Via Gobetti 93/2, I-40129 Bologna, Italy
\and
INAF-Osservatorio di Astrofisica e Scienza dello Spazio di Bologna,
Via Piero Gobetti 93/3, I-40129 Bologna, Italy
\and
INFN-Sezione di Bologna,
Viale Berti Pichat 6/2, I-40127 Bologna, Italy
\and
Department of Physics, Kavli Institute for Astrophysics and Space Research, Massachusetts Institute of Technology, Cambridge, MA 02139, USA
\and 
Fachbereich Physik, Philipps Universit\"at Marburg, D-35032 Marburg, Germany
}

  \abstract
  {The clustering of dark matter halos depends not only on halo mass but also on secondary halo properties, a phenomenon known as halo assembly bias. Coupled with variations in halo occupation, this dependence impacts galaxy clustering, an effect termed galaxy assembly bias. While previous studies have found only weak cosmological dependence of assembly bias within the $\Lambda$CDM framework, its sensitivity to the underlying dark matter physics remains largely unexplored. We use the matched dark-matter-only and hydrodynamical simulations from the AIDA-TNG suite to measure assembly bias across four dark matter models: cold dark matter, warm dark matter, and self-interacting dark matter with constant cross-sections and velocity-dependent cross-sections. We find that at $z = 0$, halo assembly bias is nearly identical across the four models for concentration-selected halos, as is the concentration dependence of the halo occupation distribution. The resulting galaxy assembly bias signals likewise show only weak model dependence. For stellar-mass-selected galaxy samples with $\mathrm{n} = 0.0586 \, h^{3} \mathrm{Mpc}^{-3}$ at $z = 0$, the velocity-dependent self-interacting dark matter scenario exhibits the largest deviation from the fiducial cold dark matter model: its assembly bias contribution to galaxy clustering ($\xi/ \xi_{\rm sh}$) is enhanced by up to $\sim 6\%$, corresponding to a $\sim 3\%$ difference in the assembly bias amplitude ($b_{\rm g} = \sqrt{\xi/ \xi_{\rm sh}}$). Although statistically distinguishable in the matched simulations, this difference is small and unlikely to be detected in current analyses. The weak model dependence persists across different galaxy number densities and redshifts. Overall, within the TNG galaxy-formation framework, these results indicate that assembly bias is largely insensitive to the dark matter models explored here, providing a robust baseline for future studies of the galaxy-halo connection in alternative dark matter scenarios.}


   \maketitle
   \nolinenumbers

\section{Introduction}

In the standard picture of structure formation, dark matter halos grow hierarchically through accretion and mergers (\citealt{Press.1974}), with galaxies residing inside them and evolving alongside them (e.g., \citealt{White.1996, Wechsler.2018}). The formation and evolution of dark matter halos are primarily governed by gravitational interactions, allowing for accurate predictions through high-resolution numerical simulations. Within this underlying dark matter framework, galaxy formation introduces an additional layer of complexity, driven by the interplay between hierarchical structure formation and detailed baryonic processes. Investigating the relationship between galaxies and halos is therefore essential for advancing our understanding of galaxy formation and for constraining cosmological models.  

In recent decades, cosmological simulations have demonstrated that the clustering of halos depends on multiple secondary halo properties beyond halo mass, such as concentration, halo age, and spin, an effect commonly known as {\it halo assembly bias} (HAB; \citealt{Sheth.2004,Gao.2005, Gao.2007,Salcedo.2018,Ramakrishnan.2019,Sato-Polito.2019, Contreras.2021,Montero-Dorta.2025}). If galaxy properties correlate with halo formation history, the galaxy content of halos is also expected to depend on secondary properties at fixed halo mass, producing variations in the halo occupation function (termed {\it occupancy variations}; \citealt{zhu.2006,Zehavi.2018, Artale.2018, Bose.2019}). Furthermore, the coupling between occupancy variation and halo assembly bias imprints the assembly history of dark matter halos onto the galaxy distribution, resulting in an additional dependence in galaxy clustering beyond halo mass, referred to as {\it galaxy assembly bias} (GAB; \citealt{Croton.2007, Faltenbacher.2010, Montero.2016, Mao.2018,  Artale.2018, Zehavi.2018, Zehavi.2019, Bose.2019, Contreras.2019, Xu.2021b, Wang.2025}). Understanding HAB and accurately modeling GAB is essential for mapping the complex galaxy-halo connection and extracting unbiased cosmological constraints from large-scale structure measurements (e.g., \citealt{Contreras.2023, Chaves-Montero.2023}).

Most previous studies of HAB and GAB have been conducted within the standard $\Lambda$CDM framework, assuming fiducial cosmological parameters. Only a few studies have explored the sensitivity of assembly bias to cosmological variations. For instance, \cite{Lazeyras.2021} demonstrate that the HAB predicted by $N$-body simulations is essentially independent of the total neutrino mass. \cite{Contreras.2021} systematically examine the impact of various cosmological parameters, concluding that the dependence of HAB on cosmology is weak for both concentration- and spin-selected halos. Furthermore, they find that the cosmological dependence of GAB is also minimal, at the level of only a few percent in the total galaxy clustering signal, and is subdominant compared to the dependence on the galaxy formation model. These results suggest that assembly bias is remarkably robust to variations of cosmological parameters within the $\Lambda$CDM paradigm. However, it remains unexplored whether the assembly bias is sensitive to modifications in the underlying dark matter physics itself.

Over the years, the standard collisionless cold dark matter (CDM) paradigm has been challenged by persistent discrepancies between observations and numerical predictions on small, non-linear scales (see \citealt{Bullock.2017} for a review). Recent strong gravitational lensing measurements reveal additional tensions concerning the properties of subhalos at both galaxy scales (e.g., \citealt{Enzi.2025, Despali.2025b, Vegetti.2026}) and cluster scales (e.g., \citealt{Meneghetti.2020,Ragagnin.2022,Meneghetti.2023}). These findings have motivated alternative dark matter models such as warm dark matter (WDM) and self-interacting dark matter (SIDM). WDM assumes non-negligible initial thermal velocities of dark matter particles, which suppress the small-scale initial power spectrum through free-streaming, delaying the formation of low-mass halos and reducing the abundance of small-scale structure (e.g., \citealt{Bode.2001, Viel.2012, Paduroiu.2022}). SIDM instead introduces scattering between dark matter particles, which redistributes energy and momentum, isotropizing halo cores and altering subhalo populations (e.g., \citealt{Spergel.2000, Vogelsberger.2012, Tulin.2018, Despali.2026, Klemmer.2026}). Although these scenarios modify halo abundance and internal structure, it remains unclear whether these effects influence HAB, propagate into the galaxy-halo connection, and ultimately produce a measurable impact on GAB.

Recent studies have begun to connect assembly bias with alternative dark matter scenarios. For instance, \cite{Zhang.2025} apply a semi-analytical model to CDM simulations to explore how SIDM-induced modifications to halo cores can link galaxy diffuseness to halo assembly history. Through halo assembly bias, this connection offers a possible explanation for the unexpectedly strong large-scale clustering of isolated, diffuse, blue dwarf galaxies in the Sloan Digital Sky Survey, a signal not reproduced by the $\Lambda$CDM galaxy-formation models examined in their study. Meanwhile, \cite{Wang.2026} investigate high-redshift HAB and apply the resulting secondary-bias trends to various formation scenarios for the Little Red Dots (\citealt{Kocevski.2023, Matthee.2024, Greene.2024}). They discuss a SIDM core-collapse scenario, where candidate hosts are selected based on early formation and high concentration. However, both studies rely on CDM-based simulations or secondary prescriptions, and do not directly measure assembly bias in self-consistent hydrodynamical simulations with alternative dark matter physics.

In this work, we use the $110.7 ~\mathrm{Mpc}$ boxes of the new comprehensive AIDA-TNG suite (Alternative Dark Matter in the TNG Universe; \citealt{Despali.2025}) to systematically investigate halo and galaxy assembly bias across four distinct dark matter scenarios: CDM, WDM, SIDM with constant cross-section (hereafter SIDM), and SIDM with velocity-dependent cross-section (hereafter vSIDM). By extending the IllustrisTNG baryonic physics (\citealt{Weinberger.2017, Pillepich.2018}) to alternative dark matter models with identical cosmological parameters, matched simulation volumes and numerical resolution, AIDA-TNG delivers the first suite of high-resolution hydrodynamic simulations that provides a controlled framework to isolate the specific impact of dark matter physics on structure formation. 

Leveraging this controlled setup, we first measure HAB in the dark matter-only runs by comparing the large-scale bias of high- and low-concentration halos at a fixed halo mass. We then utilize the hydrodynamical catalogs to characterize halo occupancy variations with concentration. We finally measure GAB by comparing the clustering of the original galaxy samples against corresponding {\it shuffled} catalogs, where assembly bias is explicitly removed. We conclude by verifying the robustness of these GAB signals for different galaxy number densities and redshifts. To our knowledge, this is the first study to compare HAB, occupancy variation, and GAB across multiple alternative dark matter models using matched dark matter-only (hereafter DMO) and full-physics hydrodynamical simulations. We find that both halo and galaxy assembly bias depend only weakly on the dark matter model over the mass, scale, and redshift ranges examined, indicating that changes in halo structure do not necessarily translate into comparably strong variations in assembly-dependent clustering.

The paper is organized as follows. In Section~\ref{sec:Method}, we describe the AIDA-TNG simulation suite and the methodology used to measure halo and galaxy assembly bias. Section~\ref{sec:HAB} examines the dependence of the HAB signal for concentration-selected halos on the underlying dark matter physics. The main results for GAB are presented in Section~\ref{sec:GAB}, which analyzes the occupancy variation and the resulting GAB signal in galaxy clustering across the four alternative dark matter models, followed by an extension to different number densities and higher redshifts. We summarize our results and conclude in Section~\ref{sec:conc}. Appendix~\ref{app:peak} justifies our choice of halo-mass binning over peak-height binning for the HAB measurements, and Appendix~\ref{app:spin} presents additional results for HAB and occupancy variation using halo spin as the secondary property.

\section{Method}
\label{sec:Method}

\subsection{Simulations}
\label{subsec:AIDA}

In this work, we analyze galaxy and halo samples from the AIDA-TNG simulations (Alternative Dark Matter in the TNG Universe; \citealt{Despali.2025, Despali.2026, Romanello.2026, Giocoli.2026}), a suite of cosmological simulations that extends the IllustrisTNG project—the “Next Generation” Illustris suite of cosmological hydrodynamical simulations (\citealt{Pillepich.2018, Springel.2018, Marinacci.2018, Nelson.2019})—to alternative dark matter models within a consistent numerical framework. The simulations adopt the fiducial IllustrisTNG galaxy formation model (\citealt{Weinberger.2017, Pillepich.2018b}) while varying the underlying dark matter physics, allowing direct comparison across different dark matter scenarios, spanning cold, warm and self-interacting dark matter. The AIDA-TNG suite features three cosmological volumes with side lengths of $25.0$, $51.7$ and $110.7 ~\mathrm{Mpc}$, the latter two corresponding to TNG50 and TNG100, respectively. Each box is simulated in up to six dark matter scenarios at two different resolution levels, including configurations both with and without baryons.

We focus on the highest available resolution of the largest simulation volume, with a comoving box size of $\sim 75\,h^{-1}\mathrm{Mpc}$, which is commonly referred to as 100/A. We analyze both the DMO and full-physics realizations. In the full-physics runs, the dark matter and baryon mass resolutions are $4.1 \times 10^7 \,h^{-1}\rm M_{\odot}$ and $0.7 \times 10^7 \,h^{-1}\mathrm{M_{\odot}}$, respectively. The corresponding DMO runs have a mass resolution of $4.8 \times 10^7 \, h^{-1}\rm M_{\odot}$. The gravitational softening length is set to $\epsilon_{\mathrm{DM, *}} = 1.48 \,\mathrm{kpc}$. The simulations adopt cosmological parameters consistent with \citet{Planck.2016}, namely, $\Omega_M = 0.3089$, $\Omega_{\Lambda} = 0.6911$, $\Omega_b = 0.0486$, $\sigma_8 = 0.8159$, and $h = 0.6774$. The initial conditions were generated at $ z=127$ using the \texttt{N-GENIC} code \citep{Springel.2005} with the Zel’dovich approximation, following the setup of the IllustrisTNG simulations \citep{Pillepich.2018}. 

We base our analysis on four dark matter scenarios provided by the AIDA-TNG 100/A boxes: the standard cold dark matter (CDM) baseline, a warm dark matter (WDM) model, and two self-interacting dark matter scenarios. Among them, the CDM run is equivalent to the TNG100-2 simulation in terms of resolution and setup, providing a well-tested reference model for comparison. The WDM model assumes a thermal relic dark matter particle mass of $\rm m_{WDM} = 3\,\mathrm{keV}$. The associated free-streaming suppresses the small-scale initial power spectrum, which is modeled explicitly and then used to regenerate the WDM initial conditions with the \texttt{N-GENIC} code (we refer the reader to \citealt{Despali.2025} for detailed information). For the self-interacting dark matter model, AIDA-TNG considers two cases: a simple scenario with a constant cross-section, $\rm \sigma/m_\chi = 1\,\mathrm{cm}^2\,\mathrm{g}^{-1}$ (hereafter SIDM), and a more realistic case with a velocity-dependent cross-section based on \citeauthor{Correa.2022} (\citeyear{Correa.2022}; hereafter vSIDM), which rapidly decreases at the typical velocity dispersion of massive bound systems. The simulations adopt the self-interaction scheme implemented in \texttt{AREPO} by \citet{Vogelsberger.2012, Vogelsberger.2016, Vogelsberger.2019}. Both self-interacting runs use the same initial conditions as CDM, since self-interactions do not significantly modify the initial perturbation field and become relevant primarily during the late, non-linear stages of structure formation.

Similar to the IllustrisTNG project, AIDA-TNG was performed with the quasi-Lagrangian \texttt{AREPO} code \citep{Springel.2010} to follow the coupled dynamics of dark matter and gas cells. Halos are identified using a friends-of-friends group finder algorithm, with a linking length of $0.2$ in units of the mean inter-particle separation (\citealt{Davis.1985}). Substructures within each FoF group are then identified with the \texttt{SUBFIND} algorithm \citep{Springel.2001}, providing subhalo and galaxy catalogs with the same set of properties available for the IllustrisTNG simulations \citep{Nelson.2019}.

To measure the halo assembly bias signal, we focus on the halo sample from the DMO runs at $z = 0$. To compute the linear halo bias, we use diluted samples of dark matter particles as the underlying dark matter density field, which is constructed by randomly selecting $0.5\%$ of the particles in the snapshot. We have tested alternative sampling fractions of $0.05\%$ and $1\%$, finding consistent results. For the galaxy assembly bias analysis, we use galaxy samples from the full-physics AIDA-TNG simulations across a range of redshifts: $z = 0, 1, 2$.

\subsection{Measuring halo assembly bias}
\label{subsec:bias}

To characterize halo assembly bias in alternative dark matter models, we measure the large-scale halo bias as a function of halo mass for both the full halo population and for sub-samples selected by certain halo properties. Note that we select halos by halo mass rather than peak height, as is commonly adopted in previous HAB studies (e.g., \citealt{Mo.1996,Contreras.2021}). Although peak height provides a useful proxy for linear halo rarity across varying redshifts and cosmologies, its mapping to the nonlinear halo abundance is not universal for models with suppressed small-scale power spectrum. We therefore adopt fixed halo-mass bins for our $z = 0$ HAB analysis to enable a direct and consistent comparison among the four models while avoiding a model-dependent peak-height prescription (see Appendix~\ref{app:peak} for further discussion and robustness tests). Specifically, we divide the halos into $0.1$ dex bins over $10.0 < \log(\mathrm{M_h}/h^{-1} \mathrm{M_{\odot}})<11.0$, $0.2$ dex bins over $11< \log(\mathrm{M_h}/h^{-1} \mathrm{M_{\odot}})<12$, and $0.5$ dex bins at the massive end. Here, $\rm M_h \equiv M_{200c}$ is the mass enclosed within $\rm R_{200c}$, where the mean enclosed density is $200$ times the critical density of the Universe $\rm \rho_c$. This binning scheme provides finer mass resolution for the abundant low-mass halo population while retaining sufficient statistics at higher masses to ensure robust clustering measurements for rarer, massive halos.

In this work, we consider halo concentration and halo spin as secondary halo properties, focusing on concentration in the main text and presenting the spin-based analysis in Appendix~\ref{app:spin}. Within each halo mass bin, we rank the halos by their secondary property and identify those in the top and bottom $20\%$ of the distribution. While alternative dark matter physics may induce shifts in absolute halo concentration \citep{Despali.2026}, defining sub-samples by percentile thresholds naturally absorbs any constant or proportional offsets across models. Using the \texttt{Corrfunc} package \citep{Sinha&Garrison.2020}, we then measure the halo-matter cross-correlation functions of the full sample and the two concentration-selected sub-samples using the diluted dark matter density field described in Section~\ref{subsec:AIDA}. The scale-dependent halo bias is computed as the ratio of the halo-matter cross-correlation function, $\rm \xi_{\rm hm}(M_h, r)$, to the auto-correlation function of the dark matter density field, $\rm \xi_{\rm mm}(r)$, 
\begin{equation}
\rm b (M_h,r) = \frac{\xi_{\rm hm}(M_h, r)}{\xi_{\rm mm}(r)},
\end{equation} 
We use the halo-matter cross-correlation rather than the halo auto-correlation to reduce statistical noise, particularly for halo sub-samples containing few objects. To minimize contributions from nonlinear clustering while retaining an adequate signal-to-noise ratio, we average this ratio over scales \(0.5 < \log(\mathrm{r}/h^{-1}{\rm Mpc}) < 0.9\) to obtain a robust linear bias value for each mass bin. This range captures the approximately scale-independent regime, circumventing the increased statistical noise at larger separations. Halo assembly bias is then evaluated by comparing this averaged bias of the high- and low-concentration subsets. Finally, performing these measurements across the CDM, WDM, SIDM, and vSIDM simulations allows us to systematically investigate how alternative dark matter physics shapes the HAB. 

We estimate the statistical uncertainties of our measurements using a jackknife resampling technique. To do so, we divide the full simulation volume into $27$ separate cubic sub-volumes by slicing each spatial axis into $3$ equal parts. Each jackknife realization consists of the full simulation box with one of the individual sub-volumes excluded. To avoid the artificial edge effects around the excluded region, for each realization, the $\rm \xi_{mm}$ is explicitly computed as the cross-correlation between the dark matter particles in the jackknife sub-sample and the full dark matter density field of the simulation, instead of the auto-correlation function of the particles in the jackknife sub-sample. Similarly, the halo-matter cross-correlation is evaluated by correlating the halos retained in the jackknife sub-sample against the full-box dark matter density field. We repeat the analysis and measure HAB for all realizations, each of which captures the sample variance contributed by the corresponding excluded sub-volume. The final error bars are then derived from the total dispersion across all jackknife realizations (see, e.g., \citealt{Lupton.1993, Zehavi.2002, Norberg.2009}). We have verified that our results are insensitive to the specific number of sub-volumes chosen.

\subsection{Measuring galaxy assembly bias} 
\label{subsec:shuff}

\begin{figure}
\centering
    \includegraphics[width=0.9\columnwidth]{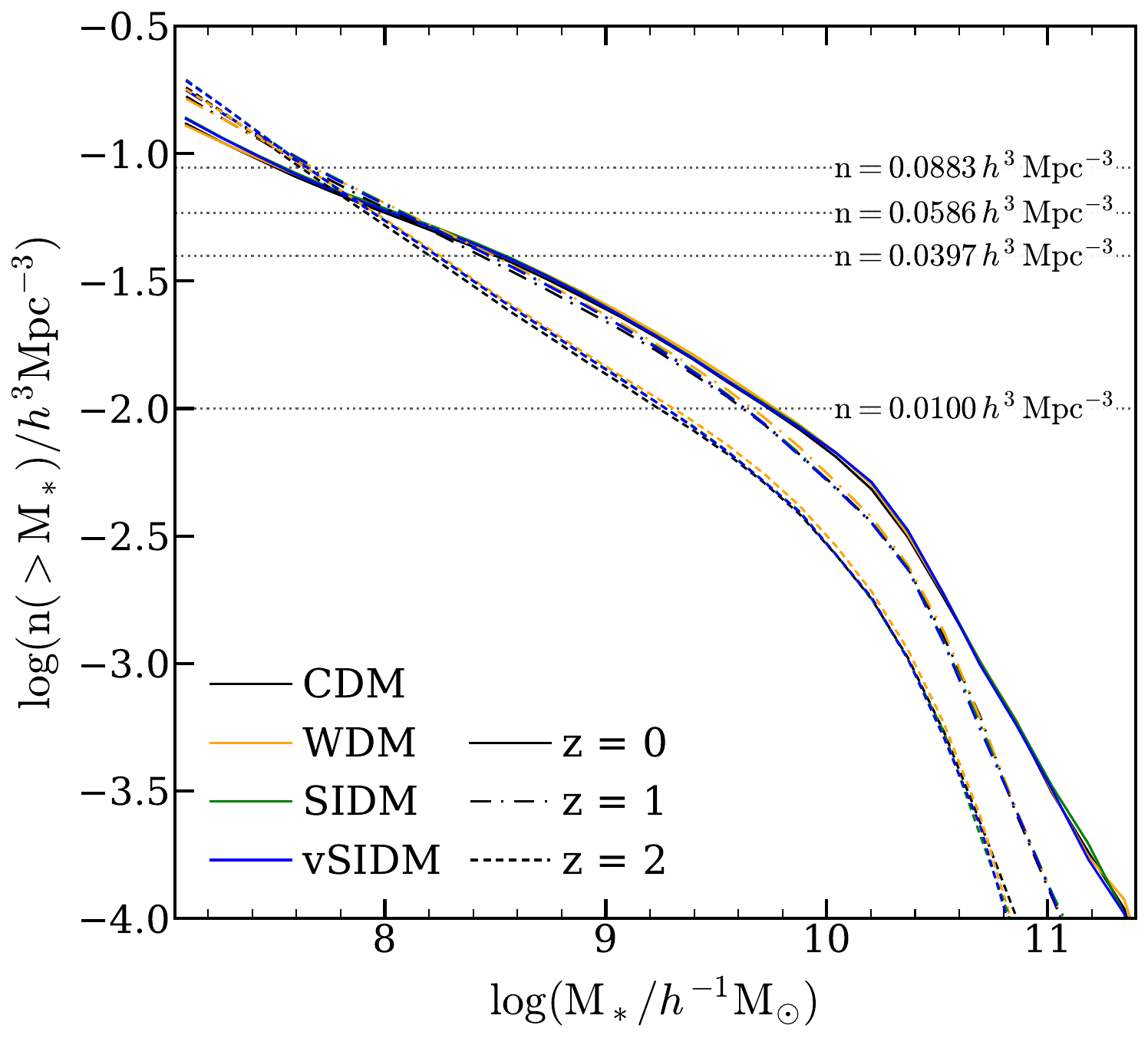}
    \caption{Cumulative stellar mass functions for the CDM, WDM, SIDM, and vSIDM galaxy sample at $z = 0, 1, 2$. Different line styles indicate the cumulative stellar mass function for galaxy samples at different redshifts, as labeled. Line colors distinguish the four dark matter models. The horizontal grey dotted lines mark the four number densities adopted in this analysis. Their corresponding values, labeled on the right-hand side of the figure, are $\mathrm{n} =0.0883, 0.0586, 0.0397, 0.0100 \, h^3\mathrm{Mpc}^{-3}$ from top to bottom.}
    \label{fig:CSMF}
\end{figure}

For the analysis at the galaxy level, we use stellar-mass-selected galaxy samples with fixed number densities rather than applying a common stellar mass threshold across all models. This guarantees identical sample abundances across all dark matter models at each redshift, allowing a direct comparison of their clustering and GAB signals. Figure~\ref{fig:CSMF} shows the cumulative stellar mass functions at different redshifts. The horizontal dotted lines indicate the four cumulative number densities adopted to define our galaxy samples, namely $\mathrm{n} =0.0883, 0.0586, 0.0397, 0.0100 \,h^3\mathrm{Mpc}^{-3}$, which correspond to minimum stellar mass thresholds of $\mathrm{M_{*, CDM}} = 10^{7.5}, 10^8, 10^{8.5},  10^{9.7} \,h^{-1} \mathrm{M}_{\odot}$ in the CDM run at $\rm z = 0$. Specifically, we choose $\mathrm{n} = 0.0586 \,h^3\mathrm{Mpc}^{-3}$ as our fiducial number density to optimize the measurements: it encompasses faint galaxies, which are expected to yield a stronger GAB signal (\citealt{Croton.2007}), and simultaneously maximizes the sample size to suppress statistical uncertainties arising from the finite AIDA-TNG volume, while remaining above the simulation's resolution limit. The other three number densities serve to test the robustness of our results against sample selection: the highest number density incorporates galaxies down to the lowest resolved stellar masses, whereas the lowest one restricts the sample to massive galaxies while maintaining sufficient statistics. The four models have nearly identical stellar mass functions at the same redshift. This is consistent with the finding in \citet{Despali.2025} that variations in the stellar mass function due to alternative dark matter models are much smaller than those produced by a modification of the baryonic treatment. 

To measure the effects of assembly bias on galaxy clustering, we construct control galaxy samples without any assembly bias and then compare them to the clustering of the original samples. To this end, we employ the standard ``shuffling method'', following the procedure of \cite{Croton.2007}. We randomly reassign the galaxy content of the halos among all halos of similar mass, thereby preserving the dependence on halo mass while removing any dependence on secondary properties such as the halo formation history, concentration, or large-scale environment. Specifically, we select halos in $0.1$ dex bins of halo mass and randomly shuffle their central galaxies among all halos within the same mass bin. Satellite galaxies are moved together with their corresponding central galaxy, preserving their relative spatial distribution within each halo. This procedure removes the dependence of halo occupation on secondary properties and their correlation with halo assembly bias, and therefore eliminates the galaxy assembly bias signal by construction.

Following standard practice, we quantify the strength of galaxy assembly bias by comparing the clustering of the original and shuffled galaxy samples. Specifically, we compute the ratio of the two-point auto-correlation functions, 
\begin{equation}
b_{\rm g}^2(r) = \frac{\xi(r)}{\xi_{\rm sh}(r)},
\end{equation}
where $\xi(r)$ and $\xi_{\rm sh}(r)$ denote the auto-correlation functions of the original and shuffled samples, respectively. Deviations of this ratio from unity on large scales quantify the contribution of assembly bias to galaxy clustering, and correspond to the square of the assembly bias signals by definition.

We have verified that our results are insensitive to the specific choice of halo mass binning, with smaller bin sizes producing negligible differences. To reduce the stochastic noise inherent in the random shuffling procedure, we generate 200 independent shuffled catalogs for each original galaxy sample, using varying random seeds. We measure the clustering signal for each shuffled catalog and adopt the average over these 200 realizations as $\rm \xi_{sh}$ throughout this work.  We have explicitly verified the convergence of the resulting GAB signal with respect to the number of realizations, finding that the mean signal stabilizes and changes negligibly once $\sim 200$ random shuffles are utilized. The uncertainties are again estimated with jackknife resampling, using the same 27 sub-volumes described in Section~\ref{subsec:bias}. For each jackknife realization, we measure galaxy clustering through the cross-correlation between the jackknife sample and the full galaxy sample, for both the original and shuffled catalogs. The jackknife uncertainty is then estimated from the scatter among the $27$ signals. Again, we have verified that our results are insensitive to the number of sub-volumes used. The variance among different shuffling realizations is negligible compared to the jackknife uncertainty, and therefore, we present only the jackknife errors in the results below.

\section{Halo assembly bias in alternative dark matter models}
\label{sec:HAB}

In this section, we examine the dependence of halo assembly bias on the underlying dark matter model using the DMO runs of the AIDA-TNG simulations. Specifically, we focus on the HAB for concentration-selected halos in the CDM, WDM, SIDM, and vSIDM models at redshift $z = 0$. Since concentration is closely connected to halo formation history and internal structure (\citealt{Giocoli.2010, Giocoli.2012}), it is one of the most well-studied secondary halo properties for characterizing halo assembly bias (e.g., \citealt{Sheth.2004, Wechsler.2006, Gao.2007, Jing.2007, Faltenbacher.2010}) and is also used in generating mock galaxy samples \citep{Hearin.2016}. In alternative dark matter models, changes in the small-scale matter distribution arising from free-streaming suppression or dark matter self-interactions can potentially alter halo structure and formation histories \citep{Lovell.2024,Despali.2026}, thereby modifying concentration-dependent halo assembly bias. For a Navarro-Frenk-White (NFW) density profile, halo concentration is conventionally defined as $\rm c_{NFW} = R_{\mathrm{200c}}/R_{\mathrm{s}}$, where $\rm R_{\mathrm{200c}}$ is the virial radius of the halo and $\rm R_s$ is the scale radius \citep{Navarro.1997}. In this work, we instead use the common proxy to characterize the concentration of each DMO halo as the ratio between the maximum circular velocity of the halo, $\rm V_{\mathrm{max}}$, and the circular velocity at the virial radius, $\rm V_{\mathrm{200c}}$ (\citealt{Bullock.2001,Prada.2012}): 
\begin{equation}
\rm c_V = V_{\mathrm{max}}/V_{\mathrm{200c}},
\end{equation}
This velocity-based proxy can be computed directly from the halo catalog and is therefore particularly useful when a robust fit to the NFW halo profile is unavailable.

\begin{figure}
\centering
    \includegraphics[width=\columnwidth]{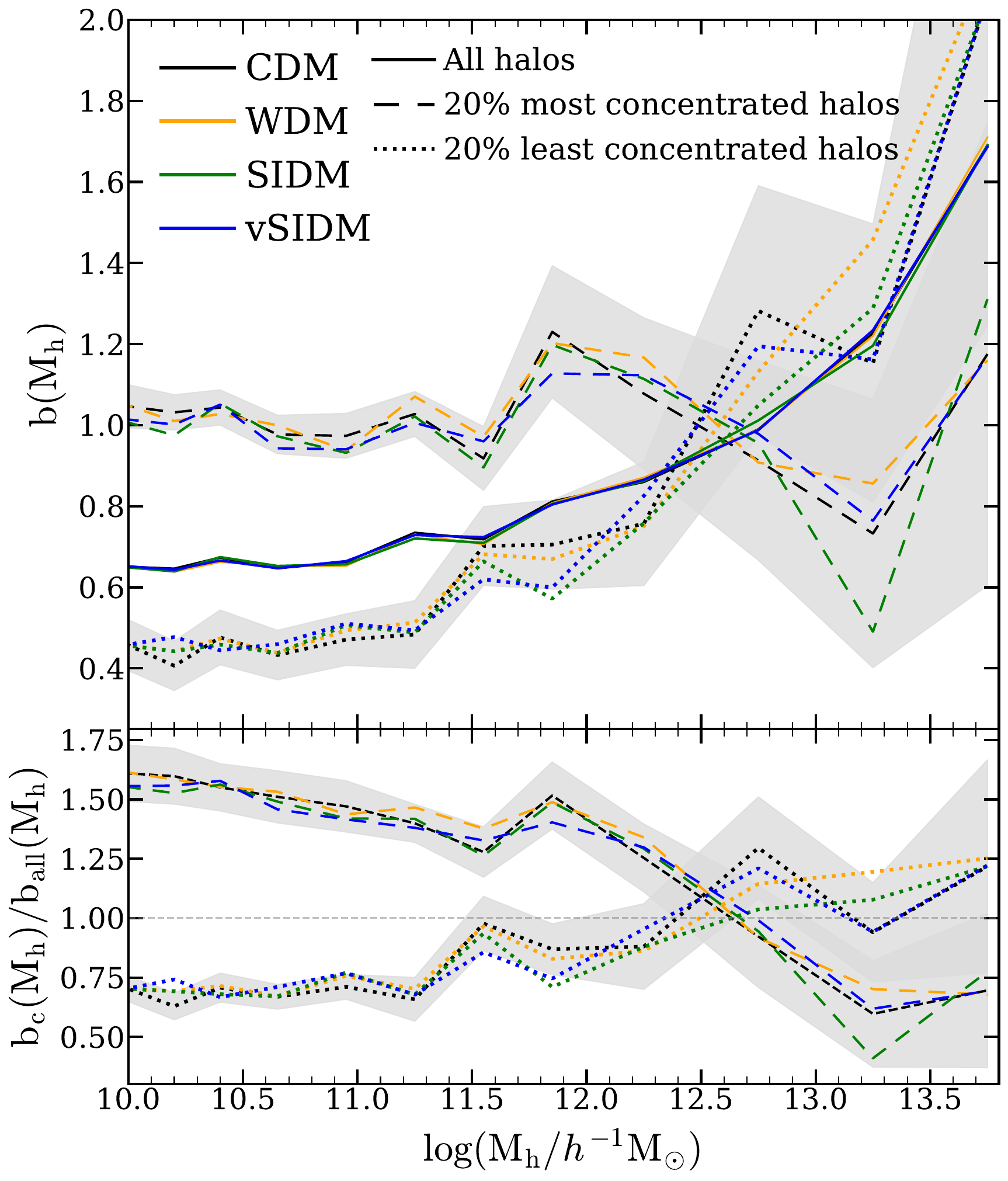}
    \caption{Halo assembly bias based on halo concentration at $z = 0$ in the DMO simulations across alternative dark matter models. The top panel shows the linear halo bias as a function of halo mass. Solid, dashed, and dotted lines denote the full sample and 20\% of halos with the highest and lowest concentration at fixed halo mass, respectively. Line colors distinguish the four simulated models. The bottom panel shows the ratio of the linear halo bias of the concentration-selected sample to that of the full sample, for the four simulations. In all panels, the shaded regions represent the error bars estimated from 27 jackknife realizations of the CDM sample, which is representative of all four models.}
    \label{fig: HAB}
\end{figure}

Figure~\ref{fig: HAB} shows the dependence of HAB on the underlying dark matter model for halos selected according to concentration. The top panel displays the linear halo bias as a function of halo mass for the four simulations, as indicated by different colors. Dashed and dotted lines represent the linear bias of the 20\% of halos with the highest and lowest concentration at fixed halo mass, respectively. The bottom panel highlights the HAB signal by showing the ratio between the concentration-selected linear bias and the full signal, $\rm b_c(M_h)/b_{all}(M_h)$. The shaded regions in each panel indicate the representative statistical uncertainty estimated from 27 jackknife realizations of the CDM sample (see Section~\ref{subsec:bias}). Since the jackknife uncertainties are comparable across the four models, we show only the CDM uncertainty for visual clarity.

Comparing the dashed lines and dotted lines in the top panel, high-concentration halos exhibit stronger clustering than low-concentration halos at fixed halo mass for halos with $\log{(\mathrm{M_h}/h^{-1}\mathrm{M_{\odot}}}) < 12.5$, while this trend weakens toward the high-mass end and reverses for $\log{(\mathrm{M_h}/h^{-1}\mathrm{M_{\odot}}}) > 12.5$. This behavior is consistent with previous studies on concentration-dependent halo assembly bias (e.g., \citealt{Wechsler.2006, Gao.2007, Contreras.2019,Montero-Dorta.2025, Wang.2026}). The same qualitative trend appears for all four dark matter models. This is more explicitly illustrated in the bottom panel, where the ratio $\rm b_c(M_h)/b_{all}(M_h)$ shows a clear separation between the high- and low-concentration sub-samples. For more concentrated halos, we find a positive HAB signal of $\sim 50\%$ at the low-mass end, and a negative HAB signal of up to $\sim 25\%$ in the high-mass regime.

More importantly, the overall HAB signal is remarkably similar across the CDM, WDM, SIDM, and vSIDM models. At fixed halo mass, the linear bias of the full sample is roughly identical among the four scenarios. For the concentration-split sub-samples, any observed discrepancies are minimal and heavily dominated by statistical noise, with no clear trend. As indicated by the shaded regions, statistical uncertainties are substantial, especially toward the massive end. This is mainly due to the limited volume of the AIDA-TNG box and the rapidly decreasing number of halos at high masses, which makes the bias measurements increasingly noisy. As a result, any differences among the four dark matter models are negligible within the jackknife error bars. 

Although alternative dark matter models can significantly modify internal halo structure (\citealt{Despali.2026}) and produce pronounced differences in small-scale halo and subhalo clustering (\citealt{Romanello.2026}), the large-scale linear bias and its concentration dependence remain unchanged. This suggests that the small-scale structural changes induced by dark matter microphysics are only weakly coupled to the relation between halo concentration and the large-scale environment at fixed halo mass. Complementary measurements based on halo spin are presented in Appendix~\ref{app:spin}, where we show that the weak model dependence of HAB is not specific to concentration.

\section{Galaxy assembly bias in alternative dark matter models}
\label{sec:GAB}

Concentration-selected HAB captures only one aspect of secondary halo bias, whereas GAB reflects the combined effects of all secondary dependencies in halo clustering and galaxy occupation. Therefore, in this section, we utilize the hydrodynamical AIDA-TNG simulations to investigate how alternative dark matter models impact the resulting assembly bias signature in galaxy clustering across varying number densities and redshifts.

We begin by examining occupancy variation, which describes how the halo occupation distribution depends on secondary halo properties (e.g., \citealt{Zehavi.2018, Artale.2018, Contreras.2019, Garc.2026}). This provides a direct test of whether underlying dark matter physics changes the connection between galaxies and halos before measuring the full clustering-based GAB signal. To make a direct connection with the HAB analysis in Section~\ref{sec:HAB}, we again use halo concentration as the secondary property, measured here in the full-physics simulations. Specifically, we compute the mean halo occupation functions for sub-samples split by concentration for the CDM, WDM, SIDM, and vSIDM models, using stellar-mass-selected galaxy samples with $n = 0.0586 \, h^{3} \mathrm{Mpc}^{-3}$ at $z=0$. As described in Section~\ref{subsec:shuff}, this number density is adopted as our fiducial number density and corresponds to a mass threshold of $\mathrm{M_{*,\mathrm{CDM}}} = 10^8 \, h^{-1}\mathrm{M}_{\odot}$ in the CDM run.   

\begin{figure}
\centering
    \includegraphics[width=0.9\columnwidth]{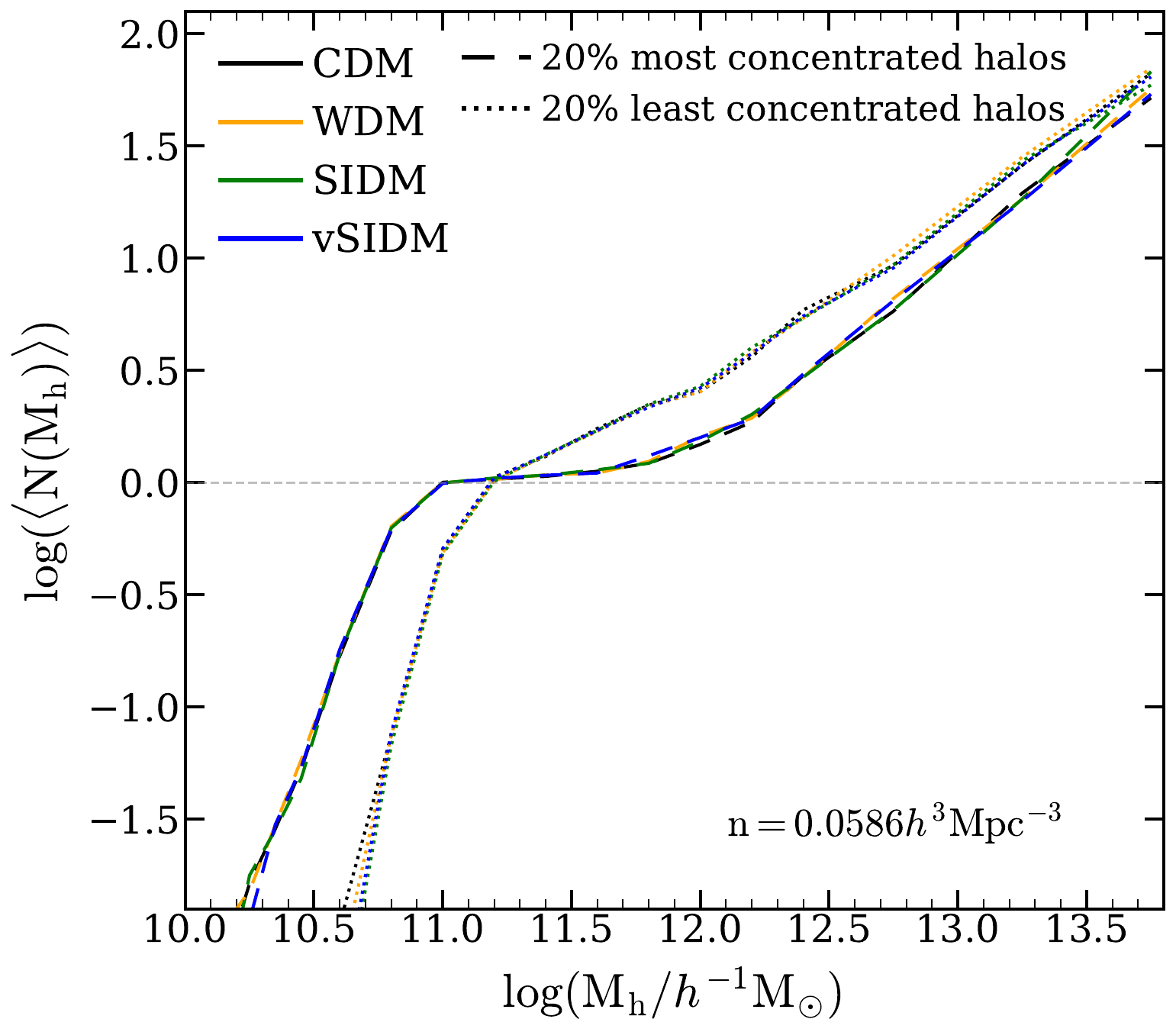}
    \caption{Concentration-dependent halo occupation functions for the CDM, WDM, SIDM, and vSIDM galaxy samples with number density $0.0586 \, h^{3} \mathrm{Mpc}^{-3}$ at $z = 0$. Different line styles denote the HOD for galaxies in the 20\% of most/least concentrated halos, as labeled. Line colors distinguish the four simulated models.}
    \label{fig:OV}
\end{figure}

Figure~\ref{fig:OV} compares the halo occupation functions of halos with the 20\% highest and lowest concentration (as indicated by the labels). These subsets are selected by ranking halos according to concentration within each halo mass bin and retaining the upper and lower 20\%, ensuring that the comparison is made at fixed halo mass and that different samples equally probe the full halo mass range.  As shown in the figure, more concentrated halos begin hosting central galaxies at lower halo masses than low-concentration systems, while hosting fewer galaxies at higher masses, consistent with previous studies (e.g., \citealt{Contreras.2019, Garc.2026}). Crucially, these occupancy variations are nearly identical across the four dark matter models at all mass scales. We have also examined the redshift evolution of the occupancy variation at the higher redshifts considered in this work ($z = 1, 2$). Although the occupation trends naturally evolve at higher redshifts (not shown), this evolution remains uniform across the models with no additional systematic dependence emerging at earlier times. This indicates that the underlying dark matter physics has a negligible impact on the concentration dependence of the galaxy-halo connection. Complementary measurements of spin-dependent occupancy variation, presented in Appendix~\ref{app:spin}, show a similarly weak dependence on the dark matter model.

We next investigate the full galaxy assembly bias signal. Galaxy assembly bias is quantified by comparing the clustering of the original galaxy sample to that of a shuffled counterpart in which correlations between galaxy occupation and secondary halo properties are removed (see Section~\ref{subsec:shuff}). We focus on the comparison between different dark matter models at $z = 0$, using galaxy samples defined by stellar-mass thresholds to match our fiducial number density.

\begin{figure}
\centering
    \includegraphics[width=\columnwidth]{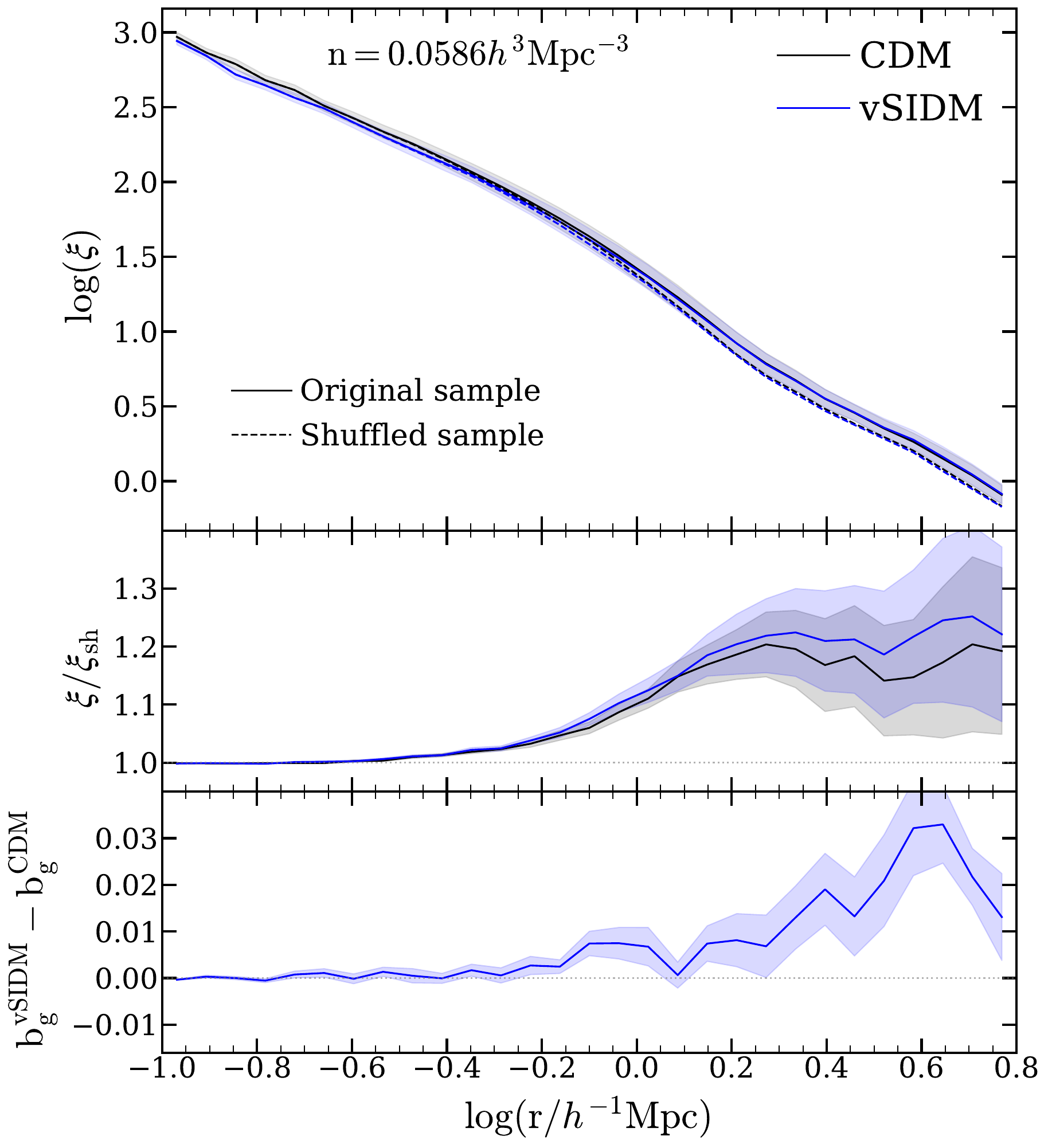}
    \caption{Two-point correlation functions for the CDM and vSIDM galaxy samples with $n = 0.0586 \, h^{3} \mathrm{Mpc}^{-3}$ at $z = 0$. The top panel shows the auto-correlation function of the original (solid line) and shuffled (dashed line) galaxy sample in the CDM simulation (black line) and the vSIDM simulation (blue line). The middle panel displays the ratio between the two-point correlation function of the original and shuffled galaxy samples (see Section~\ref{subsec:shuff}), with black and blue lines representing the CDM and vSIDM galaxies, respectively. The bottom panel shows the difference between the assembly bias signal of vSIDM and CDM, with $b_{\rm g} = \sqrt{\xi/\xi_{\mathrm{sh}}}$. In all panels, the shaded regions represent the error bars estimated from 27 jackknife realizations.}
    \label{fig:GAB-vSIDM}
\end{figure}

\begin{figure}
\centering
    \includegraphics[width=\columnwidth]{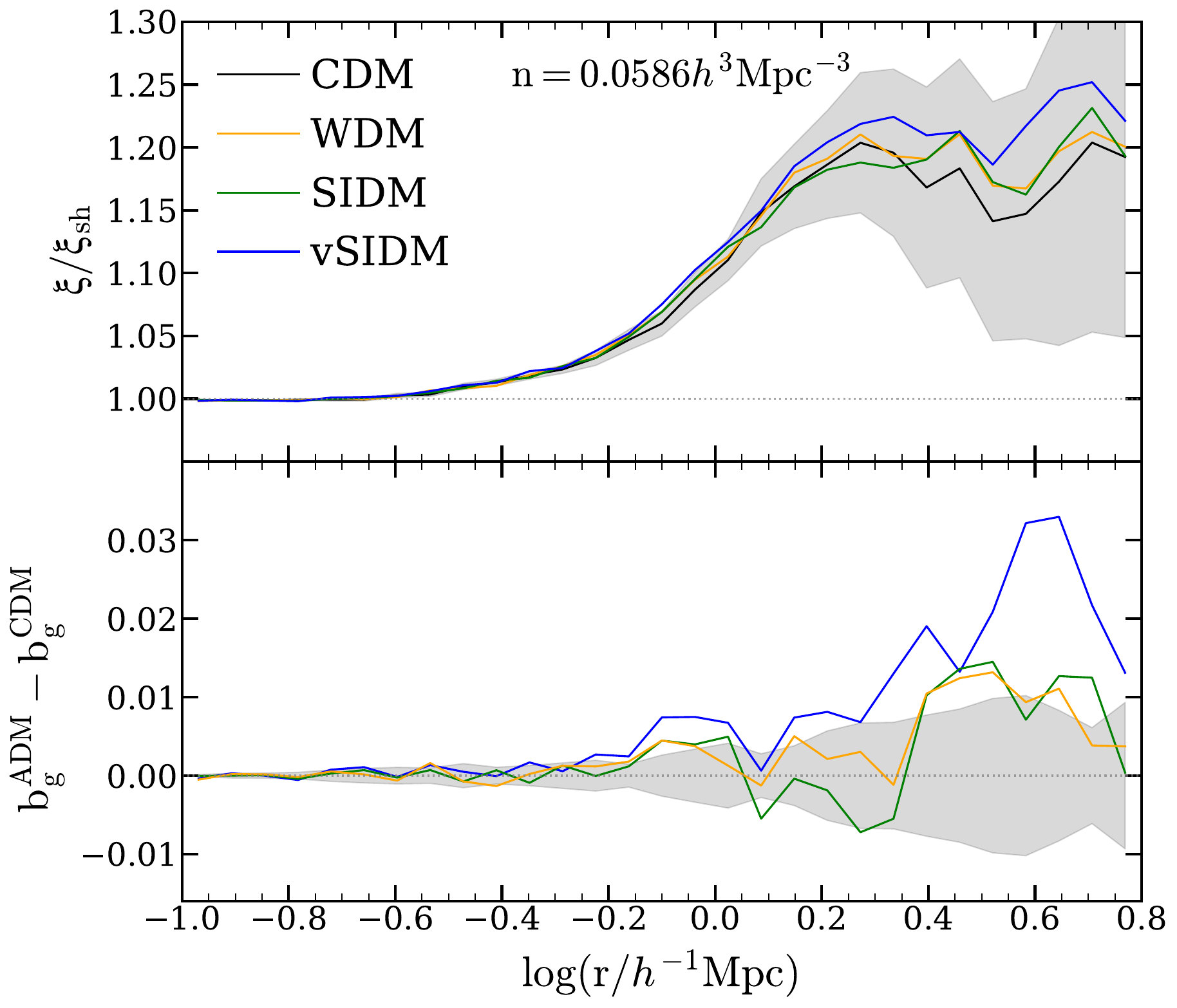}
    \caption{Galaxy assembly bias in the CDM, WDM, SIDM, and vSIDM galaxy samples with $n = 0.0586 \, h^{3} \mathrm{Mpc}^{-3}$ at $z = 0$. The top panel displays the ratio between the two-point correlation function of the original and shuffled galaxy samples (see Section~\ref{subsec:shuff}), with colors representing the galaxies in four different dark matter models as labeled. The shaded region indicates the uncertainty of the measurement in the CDM sample estimated from 27 jackknife realizations. The bottom panel shows the difference in the assembly bias, $b_{\rm g} = \sqrt{\xi/\xi_{\mathrm{sh}}}$, between each alternative dark matter model, $b_{\rm g}^{\rm ADM}$, and CDM, $b_{\rm g}^{\rm CDM}$. The shaded region shows the averaged jackknife error of the three relative model differences.}
    \label{fig:GAB-all_0}
\end{figure}

\begin{figure*}
\centering
    \includegraphics[width=\textwidth]{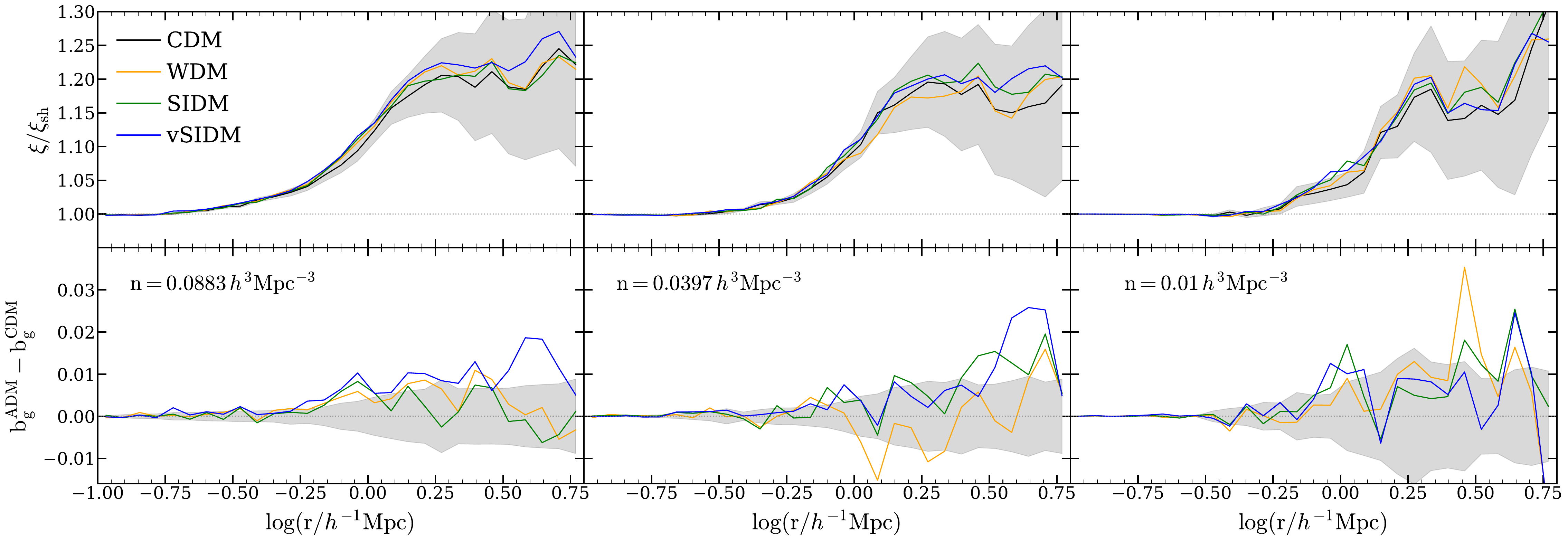}
    \caption{Galaxy assembly bias across four dark matter models with varying galaxy number densities at $z = 0$. Each column displays results for the labeled number density, which corresponds to a stellar mass threshold in the CDM sample of $\mathrm{M_{*,CDM}} = 10^{7.5} \, h^{-1}\mathrm{M_{\odot}}$ (left), $\mathrm{M_{*,CDM}} = 10^{8.5} \, h^{-1}\mathrm{M_{\odot}}$ (middle) and $\mathrm{M_{*,CDM}} = 10^{9.7} \, h^{-1}\mathrm{M_{\odot}}$ (right). All other plotting conventions follow Fig.~\ref{fig:GAB-all_0}.}
    \label{fig:GAB-3}
\end{figure*}

\begin{figure*}
\centering
    \includegraphics[width=\textwidth]{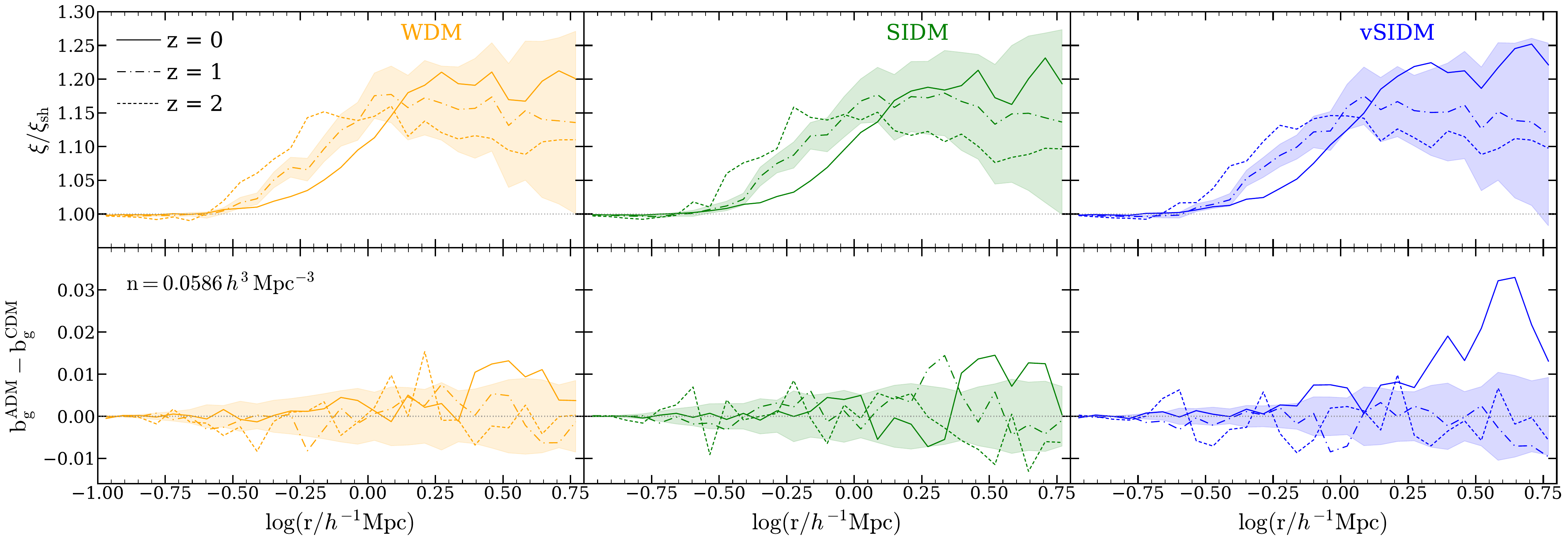}
    \caption{Redshift evolution of the galaxy assembly bias across the three alternative dark matter models. From left to right, the columns show the measurements in the WDM, SIDM, and vSIDM samples, respectively. In each panel, different line styles indicate different redshifts: $z=0$ (solid), $z=1$ (dash-dotted), and $z=2$ (dashed). All galaxy samples are selected at the same fixed number density as in Fig.~\ref{fig:GAB-all_0}, $n = 0.0586 \, h^{3} \mathrm{Mpc}^{-3}$. The bottom panels show the differences between the assembly bias signal of each alternative dark matter samples and their corresponding CDM sample at the same redshift. The shaded regions illustrate the averaged jackknife uncertainty among the three samples for each model, providing a representative estimate of the statistical errors. Other plotting conventions remain similar to those in Fig.~\ref{fig:GAB-all_0}.}
    \label{fig:GAB-z}
\end{figure*}

In order to build a clear physical picture before introducing the full four-model suite, we first focus on the CDM and vSIDM models. These two scenarios exhibit the most pronounced discrepancy, so they provide an illustrative comparison to anchor our analysis. Figure~\ref{fig:GAB-vSIDM} shows the resulting galaxy clustering and GAB measurements of galaxies in the CDM and vSIDM models. The top panel presents the two-point auto-correlation functions of the original and shuffled galaxy samples in both simulations. The middle panel shows the ratio between the original and shuffled galaxy correlation functions, $\xi/\xi_{\mathrm{sh}}$, which characterizes the assembly bias signal as described in Section~\ref{subsec:shuff}. The bottom panel further highlights the difference in the assembly bias signal ($b_{\rm g} = \sqrt{\xi/\xi_{\mathrm{sh}}}$) between the vSIDM and CDM models. The shaded regions in all panels are the uncertainties of each measurement estimated from 27 jackknife realizations, fully propagated to the signal shown. The measurements are truncated at $\log(\mathrm{r}/h^{-1} \mathrm{Mpc}) = 0.8$, as the finite simulation volume of the $75 \, h^{-1}\mathrm{Mpc}$ box leads to substantial statistical noise in the assembly bias signal on larger scales.

As shown in the top panel of Fig.~\ref{fig:GAB-vSIDM}, the auto-correlation functions of the two models exhibit subtle yet notable differences on small scales, with vSIDM galaxies clustering less strongly than CDM galaxies, before gradually converging on larger scales. This trend is qualitatively consistent with the results of \cite{Romanello.2026}, which finds that differences in halo and subhalo clustering between CDM and SIDM models are most pronounced on small scales. In both models, the shuffled samples deviate from the original samples on large scales, reflecting the contribution of galaxy assembly bias to the clustering signal. Although the overall clustering of the two models is in broad agreement, minor residual discrepancies remain, most notably in the shuffled catalogs. Specifically, comparing the blue and black curves reveals that the shuffled vSIDM galaxies cluster less strongly than the CDM ones, whereas the original samples display a milder, opposite trend. The middle panel further illustrates that the assembly-bias contribution exhibits a somewhat larger relative difference than the total correlation function. The vSIDM galaxies exhibit a slightly stronger assembly bias than the CDM model, corresponding to a $\sim 6\%$ difference in the total galaxy clustering signal on larger scales. However, due to the volume and resolution limitations, the jackknife uncertainties on the measurements are substantial compared to the GAB signals. Consequently, this noticeable discrepancy falls entirely within the statistical errors of the individual measurements. 

A more detailed comparison is shown in the bottom panel, where the relative difference between the values of $b_{\rm g}$ of the two models is measured directly. The largest deviation appears around $\log(\mathrm{r}/h^{-1} \mathrm{Mpc}) \sim 0.6$, reaching an amplitude of $\sim 3\%$. Since the CDM and vSIDM simulations share the same box and initial conditions, the sample variance partially cancels out when comparing the two models directly, allowing percent-level differences in $b_{\rm g}$ to be measured statistically, as indicated by the shaded region. Given the similar halo clustering and concentration-dependent HAB signals across the DMO runs of all four models, the small GAB difference between CDM and vSIDM likely arises from the interplay between dark matter physics and galaxy formation processes. Nevertheless, this subtle signal remains far below the sensitivity of current observational analyses. We therefore conclude that the assembly bias signal exhibits a statistically robust, yet observationally negligible difference between the CDM and vSIDM models.

Figure~\ref{fig:GAB-all_0} presents the GAB in galaxy clustering for all four dark matter models. Similar to the middle panel of Fig.~\ref{fig:GAB-vSIDM}, the top panel shows the ratio between the original and shuffled galaxy correlation functions, $\xi/\xi_{\mathrm{sh}}$, for the CDM, WDM, SIDM, and vSIDM models as labeled. For visual clarity, the shaded region shows the jackknife uncertainties only for the CDM sample, which are representative of the statistical errors across all models. We note that the total galaxy clustering of the WDM and SIDM samples (not shown) exhibits trends nearly identical to those presented in the top panel of Fig.~\ref{fig:GAB-vSIDM}. Overall, the four models exhibit very similar GAB signals across all probed scales. On larger scales, the WDM and SIDM samples follow similar trends, displaying a slightly stronger assembly bias than the CDM baseline, yet weaker than that of the vSIDM model.

The bottom panel of Fig.~\ref{fig:GAB-all_0} shows the differences in $b_{\rm g}$ between each alternative model and CDM, similar to the bottom panel of Fig.~\ref{fig:GAB-vSIDM}. Note that the shaded region centered on zero shows the mean jackknife uncertainty across the three relative deviations, providing a representative error estimate of each measurement. The WDM and SIDM differences fluctuate close to zero over most scales, indicating that the GAB amplitudes are nearly indistinguishable from CDM. vSIDM exhibits the clearest enhancement among the alternative models and is the only case whose deviation from CDM substantially exceeds the shaded uncertainty on larger scales. Nevertheless, even the largest difference remains negligible relative to the overall GAB amplitude, indicating that GAB is only weakly sensitive to the underlying dark matter physics across the models considered here.  

The weak model dependence of the full GAB signal is consistent with the concentration-dependent trends examined above. The HAB results indicate that the relation between halo concentration and the large-scale clustering is largely preserved across the four models. Likewise, the similar occupancy variations suggest that galaxy occupation in the TNG model is regulated primarily by halo mass and baryonic processes, so that changes in internal halo structure induced by the alternative dark matter models do not efficiently propagate into the galaxy-halo connection. Although concentration captures only one channel contributing to GAB, these results suggest that the coupling between secondary halo bias and galaxy occupation is not significantly altered by the dark matter models considered here.

For completeness, we extend our analysis to samples with different galaxy number densities. Figure~\ref{fig:GAB-3} presents the GAB and the residual discrepancy signals at $ z = 0$ for samples with $\mathrm{n} = 0.0883 \, h^{3} \mathrm{Mpc}^{-3}$, $ 0.0397 \, h^{3} \mathrm{Mpc}^{-3}$ and the widely adopted $ 0.01 \, h^{3} \mathrm{Mpc}^{-3}$. These selections span the practical range of galaxy catalogs that can be robustly analyzed within the AIDA-TNG box, which correspond to minimum stellar mass thresholds of $\mathrm{M_{*,CDM}} =  10^{7.5} \, h^{-1}\mathrm{M_{\odot}} $, $ 10^{8.5} \, h^{-1}\mathrm{M_{\odot}} $ and $10^{9.7} \, h^{-1}\mathrm{M_{\odot}} $ in the CDM run, respectively (see Section~\ref{subsec:shuff}). Comparing the three top panels, the GAB signals in all four models increase slightly as the number density increases. This trend is consistent with previous studies in the CDM framework that the galaxy assembly bias is stronger for fainter (less massive) galaxies (e.g., \citealt{Croton.2007, Contreras.2019}). However, as demonstrated in the bottom panels, the discrepancies between the three alternative dark matter scenarios and the CDM baseline show no systematic dependence on number density and remain smaller than for the fiducial sample in Fig.~\ref{fig:GAB-all_0}. The slightly larger deviation seen for vSIDM in the fiducial sample is likewise not amplified by selecting either fainter or more massive galaxies. We confirm that the insensitivity of the GAB signal to the underlying dark matter physics is robust and not an artifact of the chosen stellar mass threshold.

Since galaxy assembly bias is known to evolve with redshift in the standard cosmology (e.g., \citealt{Hearin.2016,Contreras.2019, Contreras.2021b,Contreras.2021}), it is natural to question whether the GAB in alternative scenarios exhibits similar redshift evolution. Furthermore, \cite{Romanello.2026} have found that the discrepancy in subhalo clustering between WDM and CDM increases toward higher redshift. While this difference is mainly confined to small scales at low redshift, it extends into the two-halo regime at $z=2$. Therefore, it is interesting to extend our analysis to higher redshifts, to test whether the weak model dependence observed at $z=0$ persists at earlier cosmic times. 

Figure~\ref{fig:GAB-z} illustrates the redshift evolution of the GAB signals across $z = 0,1,2$ for galaxies selected at the same number density as Fig.~\ref{fig:GAB-all_0}, $\mathrm{n} = 0.0586 \, h^{3} \mathrm{Mpc}^{-3}$. Each column corresponds to one alternative dark matter model, and the line style indicates the redshifts, as labeled. As shown in the top panels of Fig.~\ref{fig:GAB-z}, the overall amplitude of the galaxy assembly bias in the three models systematically decreases as the redshift increases, accompanied by a shift of the signal to smaller scales. This behavior is broadly consistent with previous CDM studies in which GAB weakens as the coupling between HAB and occupancy variation decreases at higher redshift (\citealt{Contreras.2019, Contreras.2021b}), while the shift to smaller scales likely reflects the evolving host halo population selected at fixed number density, which may change the HOD, satellite fraction, and characteristic one-halo/two-halo transition scale. More importantly, the residual signals in the bottom panels demonstrate that the subtle discrepancies between the alternative dark matter and CDM models diminish even further at higher redshifts, with no clear systematic redshift dependence. Among all samples, only the $z=0$ vSIDM case exhibits a statistically significant deviation from CDM. Repeating the comparison at other number densities yields the same qualitative behavior.

\section{Conclusion}
\label{sec:conc}

We have utilized the AIDA-TNG cosmological simulation suites to investigate the dependence of assembly bias on the underlying dark matter model. Specifically, we compare the standard cold dark matter paradigm with three variations: a warm dark matter scenario and two self-interacting models with constant and velocity-dependent cross-section. For HAB, we measure the linear halo bias as a function of halo mass for the 20\% most and least concentrated halos in each simulation, allowing us to directly compare the concentration-dependent halo assembly bias signal across the four dark matter models. As for the galaxy assembly bias, we first investigate the occupancy variation in the halo occupation functions of concentration-selected halo sub-samples, and subsequently measure the GAB signal from the ratio between the clustering of the original galaxy samples and shuffled samples. Finally, we demonstrate the robustness of our findings by extending the analysis to varying galaxy number densities and higher redshifts. Our main results are summarized as follows:

\begin{itemize}
\renewcommand{\labelitemi}{$\bullet$}
\setlength{\itemsep}{1em}

\item We detect no significant dependence of concentration-selected HAB on the dark matter model at $z=0$. Both the linear halo bias of the full halo population and the clustering difference between high- and low-concentration halos are nearly identical across the CDM, WDM, SIDM, and vSIDM simulations. Complementary measurements based on halo spin show the same qualitative behavior (Appendix~\ref{app:spin}), indicating that this robustness is not specific to concentration. 

\item The concentration dependence of galaxy occupation is likewise nearly unchanged across the four dark matter models, indicating that this aspect of the galaxy-halo connection is largely insensitive to the underlying dark matter physics considered here. An analogous result is found for spin-selected halos (Appendix~\ref{app:spin}).

\item The galaxy assembly bias signals are similar among all four models, with only small differences between alternative dark matter models and CDM. At $z=0$, the vSIDM sample exhibits the largest deviation from CDM, with a GAB contribution to the total galaxy clustering that is stronger by $\sim 6\%$ on larger scales. In terms of the assembly-bias amplitude, $b_{\rm g}$, this corresponds to a statistically significant but small difference of up to $\sim 3\%$. The WDM and SIDM models show even weaker deviations from CDM.

\item The weak dependence of GAB on the dark matter model persists across different galaxy number densities and redshifts. Although the overall amplitude of the GAB signal changes with galaxy selection and decreases toward higher redshift, the relative differences among the four dark matter models remain negligible. No clear systematic trend with redshift is observed.

\end{itemize}

Unlike previous studies that rely on analytical prescriptions or galaxy-tagging schemes applied to CDM halos to infer alternative dark matter clustering (\citealt{Zhang.2025, Wang.2026}), our analysis directly measures assembly bias in self-consistent hydrodynamical simulations of three alternative dark matter models. Overall, our results indicate that galaxy assembly bias is largely robust to the alternative dark matter physics explored in the AIDA-TNG simulations. The robustness is already evident at the halo level and persists through the galaxy-halo connection. Despite the changes in internal halo structure induced by the different dark matter models (e.g., \citealt{Despali.2026, Romanello.2026}), concentration- and spin-selected HAB in the DMO simulations remains nearly unchanged (see also Appendix~\ref{app:spin}), suggesting that the coupling between halo assembly and the large-scale environment is largely preserved. The nearly identical occupancy variations further indicate that the remaining halo-level differences are not substantially amplified through the galaxy-halo connection within the TNG galaxy-formation framework. The stability of these two ingredients naturally leads to similar GAB signals. Although concentration and spin probe only a portion of assembly bias, the close agreement of the full GAB signals suggests that other secondary dependencies introduce negligible net model dependence within the precision of our measurements.

From an observational perspective, our results establish a useful baseline for interpreting galaxy clustering in alternative dark matter scenarios. For stellar-mass-selected samples, the weak model dependence suggests that CDM-calibrated prescriptions for GAB may provide an adequate first-order description across the models considered here, with the small vSIDM offset setting a percent-level target for future precision analyses. GAB may also serve as a cross-scale consistency test: when combined with observables that directly probe halo interiors, it can determine whether small-scale modifications induced by dark matter microphysics propagate into galaxy occupation and the large-scale galaxy distribution.  

Finally, we note that our findings are inherently tied to the TNG galaxy formation model. Because subgrid baryonic prescriptions vary significantly across different hydrodynamical simulations, and given that the amplitude of galaxy assembly bias is sensitive to the details of the galaxy-halo connection, our conclusions must be interpreted within this specific context. The specific prescription of SIDM, such as the cross section, in AIDA-TNG may also influence the amplitude of the effects. For example, models with resonantly enhanced, velocity-dependent self-scattering could produce more pronounced changes in halo internal structure over specific mass ranges (e.g., \citealt{Tulin.2013a, Tulin.2013b, Tran.2025}), although whether these translate into stronger variations in HAB and GAB remains to be tested. In addition, the finite resolution and volume of the AIDA-TNG simulations limit the range of halo masses and clustering scales that can be robustly probed. Higher-resolution simulations will be particularly valuable for extending this analysis to lower-mass halos and fainter galaxies, where the GAB and impact of alternative dark matter physics may be more pronounced, while larger volumes will enable more precise measurements on large scales. Nevertheless, the AIDA-TNG suite provides a well-controlled environment for isolating the effects of dark matter physics under a fixed baryonic framework. Future hydrodynamical simulations incorporating varied alternative dark matter models alongside diverse galaxy formation prescriptions will be essential to determine whether the weak model dependence of GAB observed here constitutes a universal theoretical prediction.

\section*{Data Availability}

The original IllustrisTNG simulations are publicly available at \url{www.tng-project.org/data} \citep{Nelson.2019}. The AIDA-TNG simulations will be publicly available in the future. Additional information about
AIDA-TNG can be found online at \url{https://gdespali.github.io/AIDA/}.

\begin{acknowledgements}
      We thank Benjamin Lehmann for insightful discussions and helpful suggestions. IZ acknowledges support from a CWRU Expanding Horizons Initiative Finish Line Fund award. SC acknowledges the support of the "Ram\'on y Cajal" fellowship (RYC2023-043783-I). SC also acknowledges the support of the "Ayudas para Atracci\'on de Investigadores con Alto Potencial" (2025/00000640) from Universidad de Sevilla. JCM acknowledges financial support from the Spanish Ministry of Science and Innovation through the PID2024-159420NB-C41 project and the "Excelencia Severo Ochoa" program (CEX2024-001441-S from MICIU AEI 10.13039/501100011033), and the European Union through the ERC Consolidator Grant program (COSMO-LYA, grant agreement 101044612). GD acknowledges ISCRA and ICSC for awarding this project access to the LEONARDO supercomputer, owned by the EuroHPC Joint Undertaking, hosted by CINECA (Italy). LM acknowledges the financial contribution from the grant ASI n. 2024-10-HH.0 "Attività scientifiche per la missione Euclid – fase E". 
\end{acknowledgements}

\bibliographystyle{aa}
\bibliography{AIDA_GAB}{}

\begin{appendix}

\onecolumn
\nolinenumbers
\section{Robustness of the halo mass binning in the halo assembly bias measurement}
\label{app:peak}

Previous studies characterizing the HAB signals usually select halos according to peak height (e.g., \citealt{Mo.1996,Contreras.2021}), $\rm \nu(M_h,z) = \delta_c(z)/\sigma(M_h,z)$, where $\rm \delta_c(z)$ is the linear over-density threshold for collapse at redshift $z$ and $ \sigma(\mathrm{M_h},z)$ refers to the variance of the linear over-density field on a sphere containing the halo mass $\rm M_h$ at redshift $z$. By definition, a peak height of unity ($\nu \sim 1$) characterizes the typical mass scale of structures collapsing at the current epoch. Consequently, regions with peak heights below or above unity represent systems that have already collapsed at earlier epochs or are destined to collapse in the future (e.g., \citealt{Mo.1996, Sheth.1999}). Notably, halos with similar peak height present similar properties independently of their mass and redshift (\citealt{Sheth.1999}); meanwhile, halos of the same mass do not necessarily correspond to the same rarity in different cosmologies and redshifts. Therefore, peak height provides a useful measure of linear peak rarity and is well used to compute halo assembly bias when comparing across different cosmologies and redshifts. 

In this work, however, the peak height selection is not straightforward to apply consistently to the full set of dark matter models. In AIDA-TNG, the WDM simulation has a modified small-scale linear power spectrum, whereas the SIDM and vSIDM runs share the same initial power spectrum as the fiducial CDM model and differ only through non-linear self-interactions during structure formation (see Section~\ref{subsec:AIDA}). As a result, halos of the same mass in CDM, SIDM, and vSIDM would have the same peak height by construction, while those in WDM would have different peak heights, even though the self-interactions may also modify their internal structure and subsequent evolution. This makes a single peak-height-based comparison across CDM, WDM, SIDM, and vSIDM less straightforward than a direct halo-mass comparison at fixed redshift and cosmology.

For this reason, we adopt halo-mass bins for our HAB analysis, as described in Section~\ref{subsec:bias}. Since the comparison is performed at $z=0$ and at fixed cosmological parameters, halo mass provides a direct and transparent basis for comparing the concentration-dependent clustering signal across CDM, WDM, SIDM, and vSIDM. This choice also follows the standard definition of halo assembly bias, namely the dependence of halo clustering on secondary properties at fixed halo mass.

\begin{figure*}
\centering
    \includegraphics[width=0.5\textwidth]{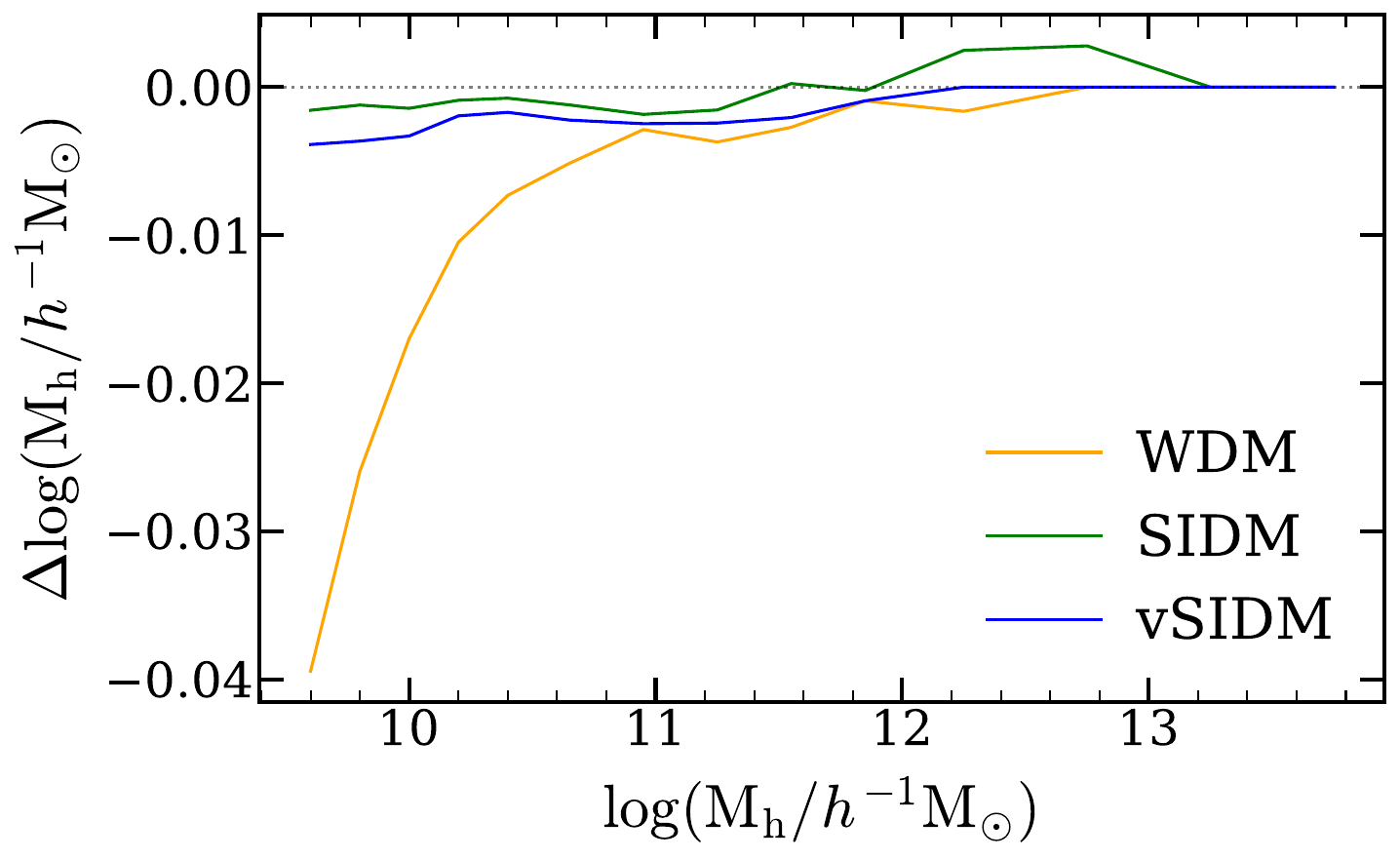}
    \caption{Mass shifts required to match the cumulative halo mass functions of the alternative dark matter models to that of CDM. Colors represent the halo samples in three different dark matter models as labeled.}
    \label{fig:abundance}
\end{figure*}

To further verify that the comparison is not affected by small differences in the halo mass functions among the models, we test an abundance-matching approach, in which halos in alternative dark matter models are matched to the CDM sample by cumulative number density. For a given CDM halo mass, we use the cumulative halo mass functions across the four simulations to identify the halo mass in each alternative dark matter model that yields the same number density. This defines a mass shift, $\Delta \log \mathrm{M_h}$, satisfying $\rm \Phi_{ADM} (\log M + \Delta \log M) = \Phi_{CDM} (\log M)$, where $\rm \Phi_{CDM}$ and $\rm \Phi_{ADM}$ denote the cumulative mass functions for the CDM and three alternative dark matter models, respectively. We then apply this mass shift to the HAB measurements, effectively re-expressing the results for alternative dark matter simulations in terms of the CDM-equivalent halo abundance. This procedure removes the impact of model-dependent differences in the halo mass function and enables a direct comparison of the concentration-dependent HAB signal at fixed statistical rarity.

As shown in Figure~\ref{fig:abundance}, the WDM model requires the largest mass shift to match the cumulative halo abundance of CDM, with a maximum absolute value of only $|\Delta \log (\mathrm{M_h}/h^{-1} \mathrm{M_{\odot}})| \sim 0.04$ at the low mass end. The corresponding shifts for SIDM and vSIDM are much smaller, with $|\Delta \log (\mathrm{M_h}/h^{-1} \mathrm{M_{\odot}})| < 0.005$. Applying these abundance-matching corrections to the halo mass produces no appreciable change in the measured HAB signals relative to our main results in Fig.~\ref{fig: HAB}. This confirms that residual differences in the halo mass functions do not drive our conclusions, and that halo-mass binning is sufficient for the $z=0$ HAB comparison presented in this work.

\section{Halo assembly bias and occupancy variation for spin-selected halos in alternative dark matter models}
\label{app:spin}

In the main text, we use concentration as the secondary property to characterize HAB and occupancy variation and find little dependence on the underlying dark matter model. To assess whether this conclusion extends beyond concentration, we repeat this analysis using halo spin as an alternative secondary property in this appendix. Halo spin is an important ingredient in galaxy formation models, as it characterizes the angular momentum of dark matter halos and is closely connected to galaxy formation and morphology (e.g., \citealt{Bullock.2001,Maller.2002, Bett.2007}). Previous studies have also identified spin as an important secondary halo property associated with halo assembly bias (e.g., \citealt{Gao.2007, Faltenbacher.2010, Lacerna.2012}). Specifically, we adopt the dimensionless spin parameter defined by \cite{Bullock.2001}, $\rm \lambda = J/(\sqrt{2} M_{200c}V_{200c}R_{200c})$, where $\rm J$ is the angular momentum within a sphere of virial radius $\rm R_{200c}$, $\rm M_{200c}$ is the corresponding virial mass and $\rm V_{200c}$ is the halo circular velocity at that radius. Following the same methodology as described in Section~\ref{sec:HAB} and Section~\ref{sec:GAB}, we rank halos by $\lambda$ within each halo-mass bin and select the 20\% with the highest and lowest spin. We then measure their large-scale halo bias in the DMO simulations and their mean galaxy occupation in the hydrodynamical simulations at $z=0$. 

\begin{figure*}
    \centering
    \makebox[0.9\textwidth][c]{
    \begin{subfigure}[t]{0.43\textwidth}
        \centering
        \includegraphics[width=\linewidth]{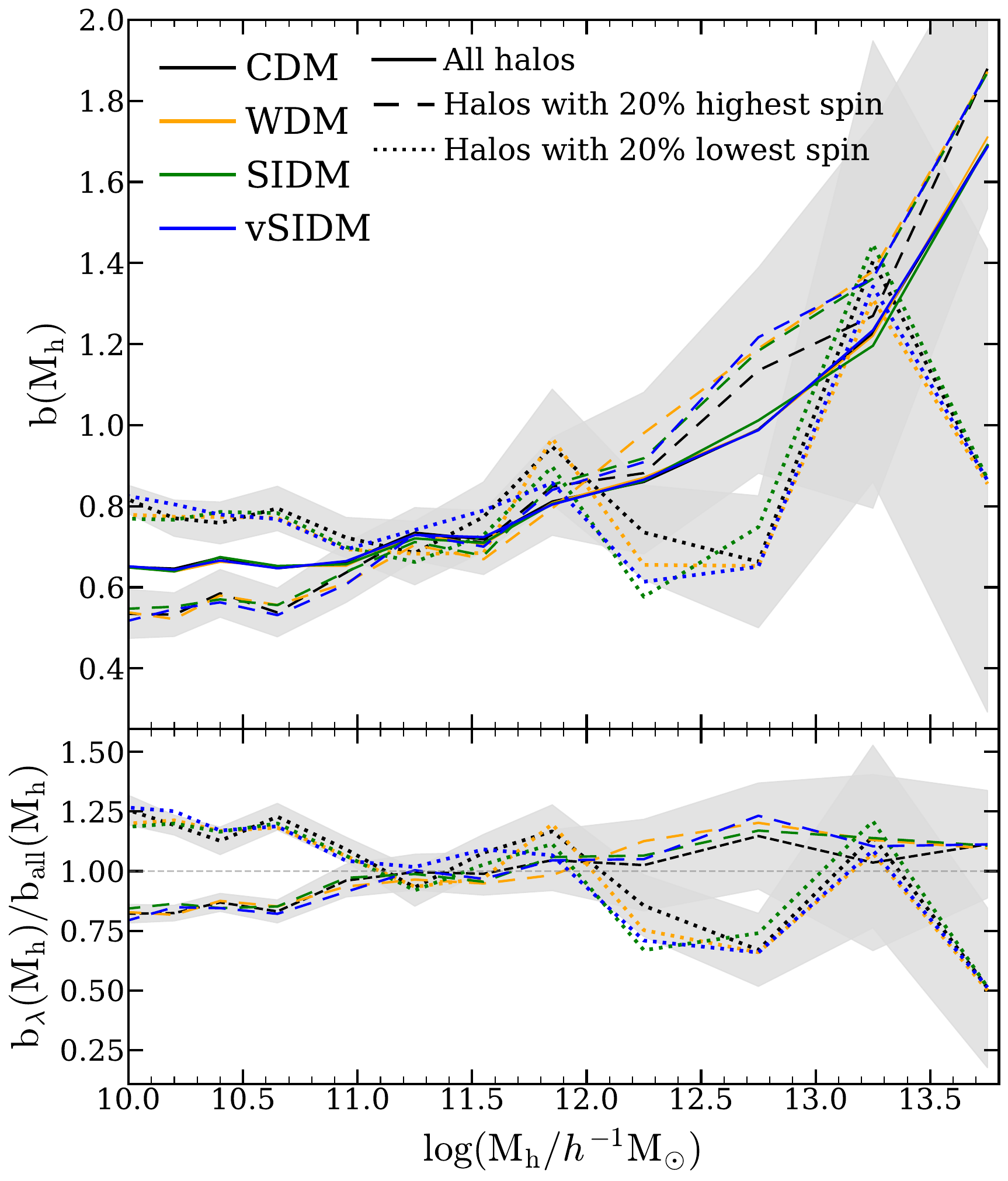}
        \caption{Halo assembly bias.}
        \label{fig:spin_hab}
    \end{subfigure}
    \hfill
    \begin{subfigure}[t]{0.43\textwidth}
        \centering
        \includegraphics[width=\linewidth]{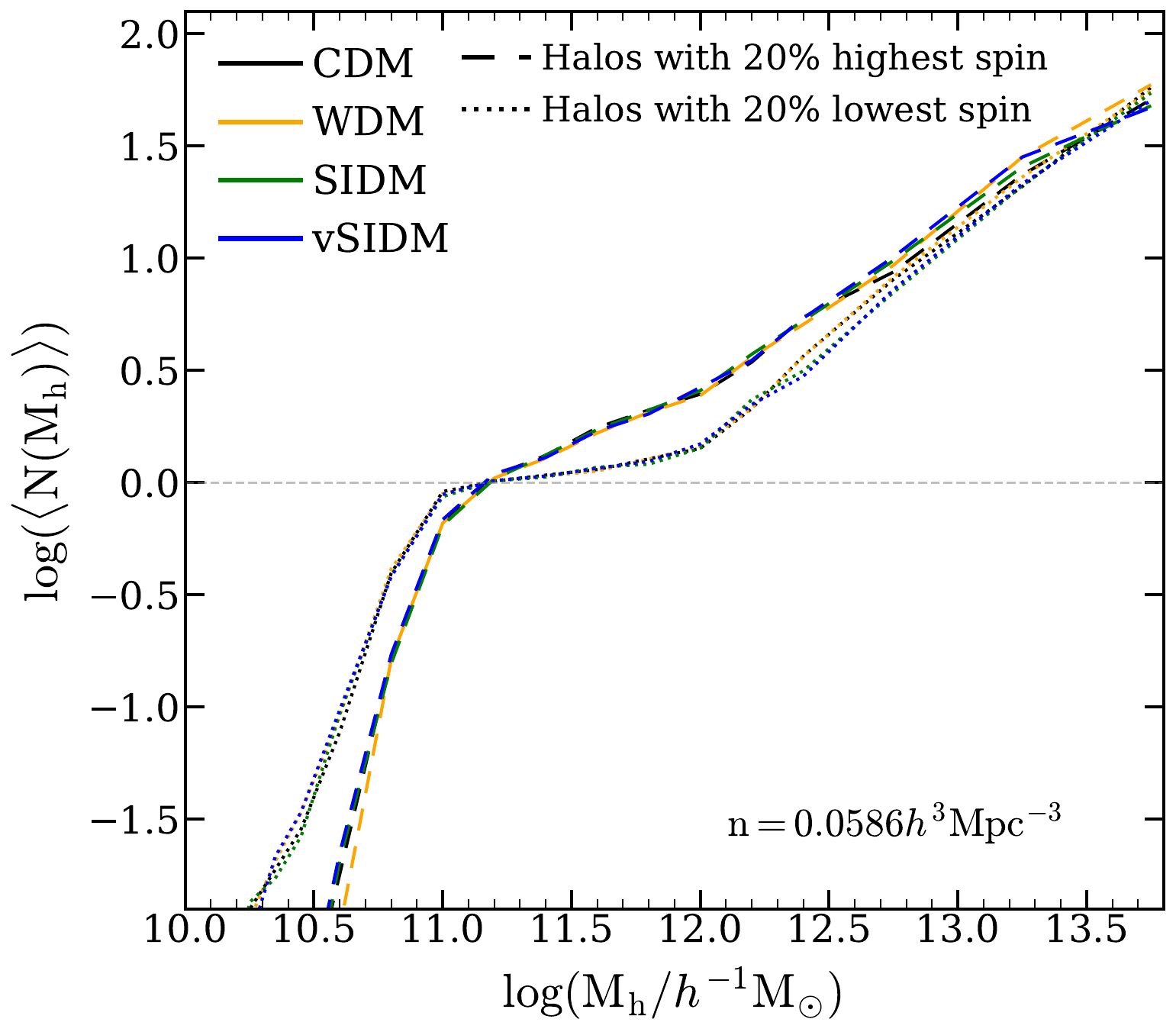}
        \caption{Halo occupation functions.}
        \label{fig:spin_occupancy}
    \end{subfigure}
}
    \caption{
        Dependence of halo assembly bias and occupancy variation on the dark matter model for spin-selected halos. Panel (a) shows the spin-selected HAB at $z=0$ in the DMO simulations, analogous to Fig.~\ref{fig: HAB}. Panel (b) shows the corresponding halo occupation functions for galaxy samples split by halo spin, analogous to Fig.~\ref{fig:OV}. Colors distinguish the four dark matter models, while line styles denote the full, high-spin, and low-spin halo samples as labeled.
    }
    \label{fig:spin_appendix}
\end{figure*}

Figure~\ref{fig:spin_hab} presents the spin-selected HAB measurements. At the low-mass end, low-spin halos are more strongly clustered than the full halo population, whereas high-spin halos exhibit a lower large-scale bias. This trend reverses at intermediate halo masses, above which high-spin halos become more strongly clustered than their low-spin counterparts. The measurements become increasingly noisy toward the massive end because of the limited number of halos in the AIDA-TNG volume. We find that the spin-selected HAB signals remain remarkably similar across CDM, WDM, SIDM, and vSIDM. No systematic separation among the four models is visible within the jackknife uncertainties, as indicated by the shaded regions.

Figure~\ref{fig:spin_occupancy} shows the corresponding spin-dependent halo occupation functions for the stellar-mass-selected galaxy samples with the fiducial number density $\mathrm{n} = 0.0586 \, h^{3} \mathrm{Mpc}^{-3}$ at $z = 0$. In the low-mass regime, low-spin halos start hosting central galaxies at lower halo mass compared to high-spin halos. At higher halo masses, the trend reverses, such that high-spin halos contain a larger mean number of satellite galaxies. These occupancy variations are nearly identical in all four dark matter models. We have also repeated the analysis at the higher redshifts considered in this work ($z = 1, 2$), finding the same negligible model dependence. Within the common TNG galaxy-formation framework, the changes in halo structure induced by alternative dark matter physics therefore do not substantially alter the spin dependence of galaxy occupation. Together, these results show that the weak model dependence found for concentration-selected HAB and occupancy variation is not specific to concentration. Although these two properties do not encompass all possible secondary dependencies contributing to GAB, their similar behavior reinforces our conclusion that assembly bias is generally insensitive to the dark matter models explored in this work.

\end{appendix}
\end{document}